\documentclass[12pt]{article}
\usepackage{graphicx,amsmath}
\usepackage{units}

\newlength{\dinwidth}
\newlength{\dinmargin}
\def\lapproxeq{\lower .7ex\hbox{$\;\stackrel{\textstyle                                                    
<}{\sim}\;$}}                                                    
\def\gapproxeq{\lower .7ex\hbox{$\;\stackrel{\textstyle                                                    
>}{\sim}\;$}}                                                    
\def\be{\begin{equation}}                                                    
\def\ee{\end{equation}}                                                    
\def\bea{\begin{eqnarray}}                                                    
\def\eea{\end{eqnarray}}

\def\bb{\vec{b}'}

\def\bb{b\bar{b}}

\def\qq{q\bar{q}}

\def\GeV{\rm GeV}

\def\sh{\hat s}
\def\sh2{{\hat s}^2}
\def\PDF{{\rm PDF}}
\def\CLO{C^{\rm LO}}
\def\CNLO{C^{\rm NLO}}
\def\CNNLO{C^{\rm NNLO}}
\begin{document}

\begin{flushright}                                                    
IPPP/12/11  \\
DCPT/12/22 \\                                                    
\today \\                                                    
\end{flushright} 

\vspace*{0.5cm}

\begin{center}
{\Large \bf Drell-Yan as a probe of small $x$ partons at the LHC}

\vspace*{1cm}
                                                   
E.G. de Oliveira$^{a,b}$, A.D. Martin$^a$ and M.G. Ryskin$^{a,c}$  \\                                                    
                                                   
\vspace*{0.5cm}                                                    
$^a$ Institute for Particle Physics Phenomenology, University of Durham, Durham, DH1 3LE \\                                                   
$^b$ Instituto de F\'{\i}sica, Universidade de S\~{a}o Paulo, C.P.
66318,05315-970 S\~{a}o Paulo, Brazil \\
$^c$ Petersburg Nuclear Physics Institute, NRC Kurchatov Institute, Gatchina, St.~Petersburg, 188300, Russia \\          
                                                    
\vspace*{1cm}                                                    
                                                    
\begin{abstract}                                                    

The predictions of Drell-Yan production of low-mass, lepton-pairs, at high rapidity at the LHC, are known to depend sensitively on the choice of factorization and renormalization scales. We show how this sensitivity can be greatly reduced by fixing the factorization scale of the LO contribution based on the known NLO matrix element,  so that observations of this process at the LHC can make direct measurements of parton distribution functions in the low $x$ domain; $x \lapproxeq 10^{-4}$. 

\end{abstract}                                                        
\vspace*{0.5cm}                                                    
                                                    
\end{center}

\section{Introduction}

The very high energy of the LHC allows us to probe the parton distribution functions (PDFs) of the proton at extremely small $x$, a region not accessible at previous accelerators.  To extract the PDFs we describe the experimentally observed cross sections as a convolution of the PDFs and the cross section for the hard subprocess. Schematically, we may write the cross section in the form
\be
\label{eq:sig}
d\sigma/d^3p~=~\int dx_1dx_2~\PDF(x_1,\mu_F)~|{\cal M}(p;\mu_F,\mu_R)|^2~\PDF(x_2,\mu_F)\ ,
\ee
where a sum over the various pairs of PDFs is implied. 
The matrix elements squared, $|{\cal M}|^2$, describe the cross sections of the elementary partonic subprocesses. An observable process at the LHC samples PDFs that carry  momenta fractions $x_i$ of that of the initial protons, where
\be
\label{eq:1}
x_{1,2}=\frac{m_{\rm hard}}{\sqrt s}\exp(\pm Y).
\ee
Here $m_{\rm hard}$ denotes the mass created in the subprocess. We see that if a relatively low-mass subsystem is observed with large rapidity, $Y$, at large collider energy, then we probe PDFs at very small values of $x_2$.
However, the problem is that using 
(\ref{eq:sig}) 
we do not know the factorization scale $\mu_F$ at which the PDFs are measured. Moreover, in the low $x$ region, the PDFs strongly depend on the choice of $\mu_F$\footnote{See the $\mu_F$ dependence of low-mass Drell-Yan production or $\bb$ production, shown, for example, in \cite{MSTWdy} and \cite{OMR} respectively.}. It is made worse due to the dominance of the
gluon PDF at small $x$.  Consider, for example, the Drell-Yan production of a low-mass,
$M$, lepton pair.  For a relatively low factorization scale the LO $q\bar{q} \to \gamma^*$ subprocess may be overshadowed by
the NLO subprocess $gq \to q\gamma^*$, due to the dominance of the gluon PDF
at low $x$.  However, as we shall see below, it is this very dominance
which will allow us to introduce a procedure which greatly suppresses
the scale dependence of the predictions.

In general, after the summation of all perturbative orders, the final result should not depend on the choice of $\mu_F$ that is used to separate the incoming PDFs from the hard matrix element.
Contributions with 
low virtuality, $q^2<\mu_F^2$, of the incoming partons are included in the PDFs, while those with 
$q^2>\mu_F^2$ are assigned to the matrix element. 
However, at low $x$ the probability to emit a new parton in some $\Delta\mu_F$ interval is enhanced by the large value of the longitudinal phase space, that is by the large value of ln$(1/x)$.
In fact, the mean number of partons in the interval $\Delta \ln \mu_F$ is
\be
\langle n \rangle \;\simeq\; \frac{\alpha_sN_C}{\pi}\; \ln (1/x)\;\Delta \ln \mu_F^2,
\label{eq:n}
\ee
leading to a value of $\langle n \rangle$ up to about 8,  for the case $\ln (1/x)\sim 8$ and the usual $\mu_F$ scale variation from $\mu /2$ to $2\mu$.  On the other hand, the NLO coefficient function (the hard matrix element) allows the emission of only {\it one} parton.  Therefore we cannot expect compensation between the contributions coming from the PDF and the coefficient function.  (At large $x$ the compensation is much more complete and provides reasonable stability of the predictions to variations of the scale $\mu_F$.)

Here, our plan is to use the NLO result to fix the proper choice of the factorization scale for the LO part of the amplitude, and to demonstrate that the resulting predictions are quite stable to variations of $\mu_F$ in the remaining NLO part.
In other words, we would like to choose the value of $\mu_F$ which minimizes the higher order $\alpha_s$ NLO, NNLO,.. contributions.

As mentioned above, we may probe PDFs at low $x$ by observing low-mass systems, such as Drell-Yan lepton-pair production or $\bb$ production, at the LHC.  In each case the subprocess mass in (\ref{eq:1}) is relatively small, and so observation at large rapidities will probe PDFs at small values of $x_2$. These kinematics are particularly optimal for the LHCb experiment \cite{LHCb}. Since Drell-Yan is the simpler example, we will use it to illustrate the method.

\begin{figure} [htb]
\begin{center}
\vspace{-1cm}
\includegraphics[height=8cm]{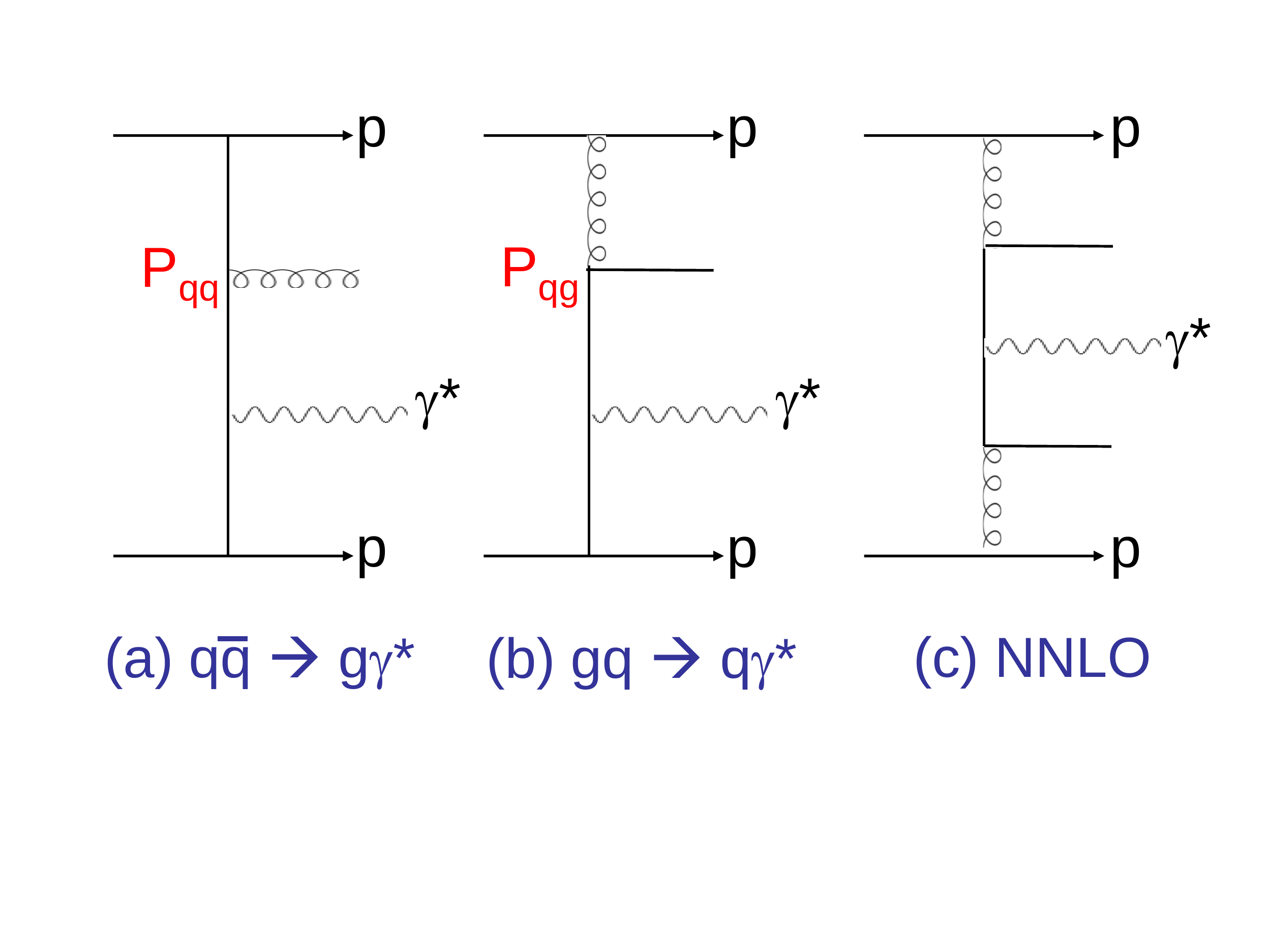}
\vspace{-2cm}
\caption{\sf Diagrams (a,b) show NLO subprocesses for Drell-Yan production resulting from the splitting of the upper PDF (or `right' PDF, in the notation of (\ref{eq:5})). Diagram (c) is the main NNLO subprocess $gg \to \bar{q}\gamma^* q$. }
\label{fig:fd}
\end{center}
\end{figure}

To sketch the idea, we start with the LO expression for the cross section. In the collinear approach, the cross section has the form
\be
\sigma(\mu_F)~=~\PDF(\mu_F)\otimes\CLO \otimes \PDF(\mu_F). 
\ee
The effect of varying the scale from $m$ to $\mu_F$, in both the left and right PDFs, can be expressed, to first order in $\alpha_s$, as
\be 
\sigma(\mu_F)=\PDF(m)\otimes    
 \left(\CLO ~+~\frac{\alpha_s}{2\pi}{\rm ln}\left(\frac{\mu_F^2}{m^2}\right)(P_{\rm left}\CLO+\CLO P_{\rm right})\right)\otimes \PDF(m),
\label{eq:5}
\ee
where the splitting functions
\begin{equation}
P_{\rm right}=P_{qq}+P_{qg}~,~~~~~~~~~P_{\rm left}=P_{\bar{q}\bar{q}}+P_{\bar{q}g}
\end{equation}
act on the right and left PDFs respectively, see Fig.~\ref{fig:fd}. We may equally well have incoming $\bar{q}$'s in $P_{\rm right}$ and incoming $q$'s in $P_{\rm left}$.
Recall that in calculating the $\alpha_s$ correction in (\ref{eq:5}),
the integral over the transverse momentum (virtuality) of the parton in the LO DGLAP evolution was approximated by the pure logarithmic $dk^2/k^2$ form. That is, in the collinear approach, the Leading Log Approximation (LLA) is used.

Now let us study the expression for the cross section at NLO. First, we note that the original Feynman integrals corresponding to the NLO matrix element, $\CNLO$, do not depend on $\mu_F$. However, we will see below how a scale dependence enters.  At NLO we may write
\be
\sigma(\mu_F)~=~\PDF(\mu_F)\otimes(\CLO + \alpha_s \CNLO_{\rm corr}) \otimes \PDF(\mu_F), 
\label{eq:stab}
\ee
where we include the NLO correction to the coefficient function.  In terms, for example, of the diagrams (a,b) of Fig. \ref{fig:fd} this means that the $2\to 2$ subprocesses $\qq \to g\gamma^*$ and  $gq \to q\gamma^* $ are now calculated with better, than LLA, accuracy. However part of this contribution was already included, to LLA accuracy, in the second term in (\ref{eq:5}).  So this part should be subtracted from $\CNLO$. Moreover, this LLA part depends on the scale $\mu_F$. As a result, changing $\mu_F$ redistributes the order $\alpha_s$ correction between the LO part ($\PDF\otimes\CLO\otimes\PDF$) and the NLO part ($\PDF\otimes\alpha_s\CNLO_{\rm rem}\otimes\PDF$).  We see the part of NLO correction which remains after the subtraction, $\CNLO_{\rm rem}(\mu_F)$, now depends on the scale $\mu_F$, coming from the $\mu_F$ dependence of the LO LLA term that has been subtracted off. The trick is to choose an appropriate scale, $\mu_F=\mu_0$, so to minimize the remaining NLO contribution $\CNLO_{\rm rem}(\mu_F)$.  To be more precise, we choose a value  $\mu_F=\mu_0$ such that as much as possible of the `real' NLO contribution (which has a `ladder-like' form and which is strongly enhanced by the large value of $\ln(1/x)$) is included in the LO part where all the logarithmically enhanced, $\alpha_s\ln(1/x)$, terms are naturally collected by the incoming parton distributions\footnote{Actually our approach is rather close to the $k_t$-factorization method. Using the known NLO result we account for the exact $k_t$ integration in the last cell, adjacent the LO `hard'
matrix element (describing heavy photon emission), while the {\it unintegrated} parton distribution is generated by the last step of the DGLAP evolution, just like the prescription proposed in \cite{KimbMR,MRW}.}.

In the next section we describe how to choose an appropriate scale $\mu_F=\mu_0$. Then in Section \ref{sec:3a} we demonstrate how by fixing $\mu_F=\mu_0$ in the LO part of (\ref{eq:5}), we obtain stability of the Drell-Yan cross section with respect to variations of $\mu_F$ in the remaining NLO contribution.  Moreover, it turns out that the NNLO correction to the Drell-Yan process at such a value of $\mu_F$ is also very small.

\section{The choice of an appropriate factorization scale \label{sec:3} }

DGLAP evolution describes the variation of parton densities $a(x,\mu)$
with the scale $\mu$. At LO it accounts for the splitting of a parton $a(x_a,k_{at})$ to a parton $b(x_b,k_{bt})$,
with the longitudinal momentum fractions which satisfy, $x_a>x_b$ and the transverse momenta which satisfy $k_{at}<k_{bt}$. This  splitting is described by  Feynman diagrams in the Leading Log Approximation (LLA) where the exact $k_t$ distribution is approximated by the pure logarithmic form $dk^2_t/k^2_t$.

However the hard matrix element is calculated more precisely. Here we account for the exact $k_t$ dependence. Our aim is to choose the upper limit $\mu_F$  in the logarithmic $k_t$ integration in such a way that the value of the integral becomes equal to the result given by the NLO matrix element which contains the exact 
$k_t$ integration. Strictly speaking this cannot be done for all the hard subprocesses simultaneously, with a common value of $\mu_F$. So we have to compromise, and to choose $\mu_F$ for the subprocess  which gives the major contribution. 

For Drell-Yan production at low $x$, the majority of quarks/antiquarks are produced via the low-$x$ gluon to quark splitting, $P_{qg}$. That is, the most important NLO subprocess is $gq\to q\gamma^* $.
The corresponding cross section reads
\begin{equation}
\frac{d\hat{\sigma}}{dt}(gq\to q\gamma^* )~ =~\frac{\pi\alpha_s\alpha^{\rm QED}}
{\hat s^2N_c}\left(-\frac{\hat s}t-\frac t{\hat s}-\frac{2M^2u}{t\hat s}\right)
\label{eq:A}
\end{equation}
where $N_c=3$ and $u=M^2-\hat s -t$. The  kinematical upper limit of $|t|$ is 
\be
|t|=\hat s-M^2=(1-z)M^2/z,
\ee
where $M$ is the mass of Drell-Yan lepton pair and\footnote{Strictly speaking, $z$ is the ratio of the light-cone momentum fraction carried by the `daughter' quark to that carried by the `parent' gluon, $z=x^+_q/x^+_g$.} $z=M^2/\hat s$.

At first sight, a problem appears to be that, for a large incoming $gq$ energy, $\sqrt{\hat{s}}$, we will have to consider very large $|t|$; that is  unreasonably large values of $k_t^2=-(1-z)t$. However, in reality this is not the 
case. Let us consider the previous splitting, $P_{gq}$, which produces the gluon from an incoming quark.
That is the $q\bar{q}\to q\bar{q}\gamma^*$ subprocess. For large energy, $s_{q\bar{q}} \gg M^2$, of the incoming $q\bar{q}$ system, the gluon flux generated by the fast quark is of the form
$dx_g/x_g$. Since the cross section  (\ref{eq:A}) decreases as the subenergy $\hat s$ increases,  weighting the $gq\to q\gamma^* $ cross section with the gluon flux, results in the main contribution coming from the relatively low $\hat s\sim M^2$ region.

An explicit calculation shows that in such a case the pure logarithmic LO
evolution integrated up to $\mu_F=\mu_0=1.4M$ convoluted with $C^{\rm LO}$ reproduces the NLO contribution of this subprocess.
Since in the low-$x$ region $gq \to q\gamma^*$ is the dominant subprocess we anticipate that the choice of $\mu_F=\mu_0=1.4M$ will minimize the remaining NLO contribution. 

Moreover, at NNLO, for very small $x$, the dominant diagram is $gg\to \bar{q}\gamma^* q$, where on both sides of $\gamma^*$ (the Drell-Yan heavy photon) the quark and the antiquark are produced via the  gluon to quark splitting, see Fig. \ref{fig:fd}(c).
That is, we have a $gg\to \bar{q}\gamma^* q$ subprocess with the same kinematics as for the NLO $qg\to q\gamma^* $ case considered above. It should be also well mimicked by the LO evolution with $\mu_F=\mu_0$. Thus the choice $\mu_F=\mu_0$ will suppress (in a low-$x$ domain) the NNLO contribution as well.

\section{Scale dependence of low-mass Drell-Yan production \label{sec:3a}}

\begin{figure} [htb]
\begin{center}
\includegraphics[height=10cm]{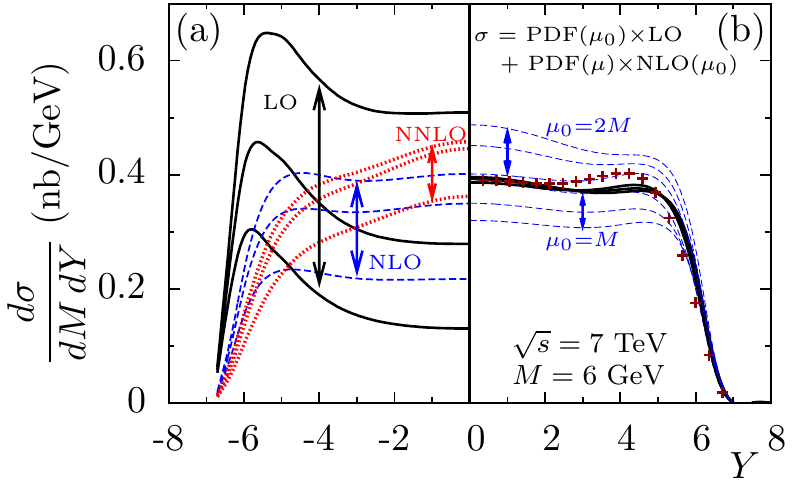}
\vspace{-0.2cm}
\caption{\sf (a) Sensitivity of $M=6$ GeV Drell-Yan $\mu^+\mu^-$ production, as a function of rapidity $Y$, to the choice of factorization scale: $\mu_F=M/2,\ M,\ 2M$, at LO, NLO, NNLO. (b) The bold lines correspond to the choice $\mu_F=\mu_0=1.4M$ which minimizes $\CNLO_{\rm rem}$, and show the stability with respect to the variations $\mu=M/2,\ M,\ 2M$ in the scale of the PDFs convoluted with $\CNLO_{\rm rem}$ -- the $\mu$ dependence is indicated by the symbolic equation at the top of the diagram. The dashed lines show that the stability disappears for other choices of $\mu_0$. The small crosses are the NNLO result. In this figure the renormalization scale is fixed at $\mu_R=M$.}
\label{fig:muF}
\end{center}
\end{figure}

\begin{figure} [htb]
\begin{center}
\includegraphics[height=10cm]{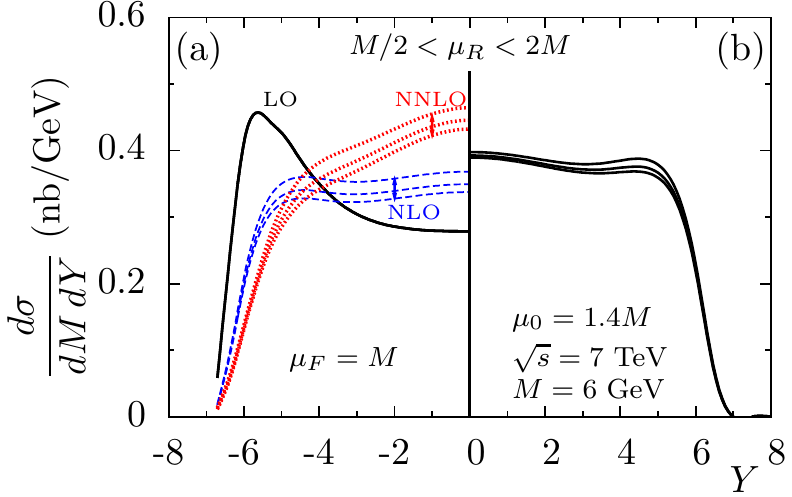}
\vspace{-0.2cm}
\caption{\sf (a) Sensitivity of $M=6$ GeV Drell-Yan $\mu^+\mu^-$ production to the choice of renormalization scale: $\mu_R=M/2,\ M,\ 2M$, at LO, NLO, NNLO; and the fixed factorization scale $\mu_F=M$. (b) Stability of the NLO result is achieved for the optimal choice $\mu_F=\mu_0=1.4M$.}
\label{fig:muR}
\end{center}
\end{figure}

Figs. \ref{fig:muF} and \ref{fig:muR} show, respectively, the factorization, $\mu_F$, and the renormalization, $\mu_R$, scale dependences of the cross section as a function of rapidity for the Drell-Yan production\footnote{We use the computer code Vrap of Ref. \cite{Vrap}.} of a $\mu^+\mu^-$ pair of mass $M=6$ GeV at the LHC energy of 7 TeV. As the plots are symmetrical about rapidity $Y=0$, we are able to compare directly two sets of distributions.   

The left-half of each of these plots 
shows the LO, NLO and NNLO results calculated with the MSTW2008 LO, NLO and the NNLO PDFs \cite{MSTW} respectively. In Fig. \ref{fig:muF}(a) the renormalization scale $\mu_R=M$ is fixed, while the value of factorization scale is varied: $\mu_F=M/2,\ M,\ 2M$. In Fig. \ref{fig:muR}(a) the factorization scale is fixed $\mu_F=M$, while the renormalization scale is varied: $\mu_R=M/2,\ M,\ 2M$. It is clear that the uncertainties arising from the $\mu_F$
dependence are huge at the LO, large at the NLO and still not negligible even at the NNLO; see, also, \cite{MSTWdy}. The dependence on the value of $\mu_R$ is not so strong, but is still not negligible.

In the right-half of these plots, the continuous bold lines show the cross sections calculated using the PDFs with our `optimal' choice of scale, $\mu_F=\mu_0=1.4M$, in the LO part. The only scale dependence in Fig. \ref{fig:muF}(b) comes from the variation of $\mu_F=M/2,\ M,\ 2M$ in the PDFs convoluted with the remaining NLO contribution. Recall that the value of $\mu_0$ was found to almost nullify the NLO correction, leaving only a small $\CNLO_{\rm rem}$ contribution. Indeed, the plot shows that the distribution has very good stability to the variation of $\mu_F$. The dashed lines on this plot (Fig. \ref{fig:muF}(b)) show that the stability would be much worse if we were to take $\mu_0=M$ or $\mu_0=2M$, rather than the optimal value of $\mu_F=\mu_0=1.4M$.

Moreover, as we argued in Section \ref{sec:3}, even the NNLO correction is rather small for the choice  $\mu_0=1.4M$. The corresponding NNLO distribution, calculated with $\mu_F=\mu_0=1.4M$, is shown by the crosses. It
is rather close to the (NLO) continuous lines. Recall, here we have used
$\mu_R=M$, and the same MSTW2008 {\bf NLO} PDFs for both the NLO and NNLO calculations, as we wish to see the effect of changing from $\CNLO$ to $\CNNLO$ leaving the PDFs unchanged.  Let us explain why this is the appropriate procedure, since at first sight it may appear that we should use NNLO PDFs for the NNLO prediction.

The basis of our method is that the long evolution length, in (\ref{eq:5}) with $m=Q_0$, means that the LLA LO result can capture most of the answer provided that the factorization scale is chosen correctly.  That is, the LO term captures most of the NLO contribution for $\mu_F=\mu_0=1.4M$, and even the major part of the NNLO contribution for the same value of $\mu_F$. 

In general, one may use the known NNLO result to determine another value of the optimal scale, $\mu_1$, such that the NLO part (the $\alpha_s$ term in the splitting functions) of DGLAP evolution will capture most of the NNLO
 contribution. However, as it is seen from Figs. \ref{fig:muF} and \ref{fig:muR}, already the 
 scale $\mu_0$ used for the LO part (the $\alpha^0_s$ term in the splitting functions) provides sufficient stability
of the results for the Drell-Yan cross section\footnote{Recall, that the subtractions of the contribution generated by the NLO DGLAP evolution convoluted with the LO matrix element from the exact NNLO matrix element include the NLO splitting functions. Therefore
we can convolute the known NNLO coefficient functions, $\CNNLO$, with either the NLO or NNLO PDFs, but not with the LO PDF; in the last case we need to recalculate the NNLO matrix element.}.

In Fig. \ref{fig:muR}(b) we see an analogous stability with respect to variations of the renormalization scale: $\mu_R=M/2,\ M,\ 2M$, when the factorization scale is fixed at $\mu_F=\mu_0=1.4M$. This additionally confirms that the NLO (and NNLO) corrections are suppressed for the optimal choice of $\mu_F$.

\section{PDF dependence  of low-mass Drell-Yan production}

What values of $x_1$ and $x_2$ of the PDFs are probed in Drell-Yan production of a system of mass $M$ with rapidity $Y$? 
Formally, in the collinear approximation, the LO term samples PDFs with precisely known momentum fractions
\begin{equation}
x_{1,2}=\frac{M}{\sqrt s}\exp(\pm Y)\ .
\label{eq:x12}
\end{equation}
That is, the $x_i$ distributions are
$\delta$-functions. This is not true in reality. Formula (\ref{eq:x12}) does not
include the transverse momenta of the incoming partons. It assumes $k_t \ll M$. If we account for $k_t$, then we would have 
\begin{equation}
x_{1,2}=\frac{\sqrt{M^2+k^2_t}}{\sqrt s}\exp(\pm Y)\ .
\label{eq:xt12}
\end{equation}
On the other hand, the optimal scale $\mu_0=1.4M$ is rather large. For this scale, the partons in the PDFs have $k_t$'s up to $\mu_0$, which may exceed the value of $M$. In order to get a more realistic impression of the interval of $x$ sampled by the process, we go back one step in the DGLAP evolution. That is, we consider the $qg\to \gamma^* q$ subprocess, which we already discussed at NLO. Recall that actually the majority of low-$x$ quarks(antiquarks) were produced just in this way. 

\begin{figure} [htb]
\begin{center}
\includegraphics[height=8cm]{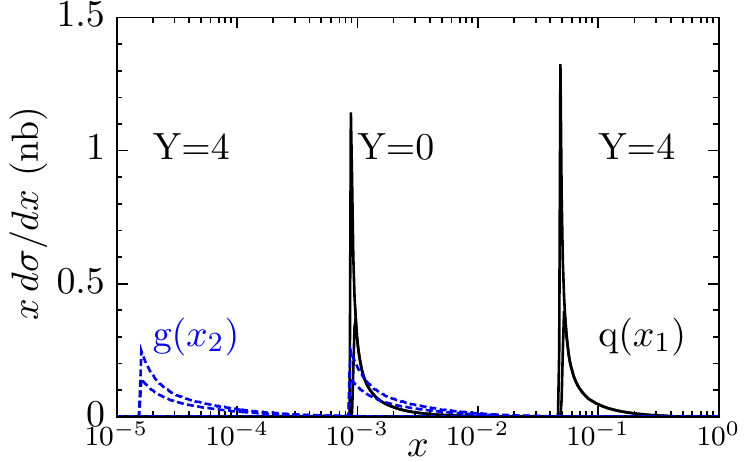}
\vspace{-0.2cm}
\caption{\sf The values of $x_1$ and $x_2$ of PDFs probed by the observation of Drell-Yan production, through the $qg$ channel, of a lepton-pair of mass $M=6$ GeV at a 7 TeV LHC. The gluon (quark) PDF is sampled in the $x$ region given by the dashed (continuous) curves. The distributions are shown for two different values of the rapidity, $Y$, of the lepton pair system. The two curves for each PDF at each rapidity correspond to two choices of the  virtuality cutoff: namely $Q_0=1$ and $Q_0=2$ GeV.}
\label{fig:hist}
\end{center}
\end{figure}
To be specific, let us take the $gq\to q\gamma^*$ matrix
 element (\ref{eq:A}), and convolute it with the NLO MSTW PDFs. Fig. \ref{fig:hist} shows the resulting distributions of the momentum fractions of the quark (continuous lines) and the gluon (dashed lines). As before we consider Drell-Yan production of a system of mass $M=6$ GeV at the LHC energy $\sqrt s=7$ TeV. We show the distributions for two rapidities: $Y=0$ and $Y=4$. The very sharp left-hand cutoff of the distributions reflects the fact that the momentum fraction cannot be less than that given by 
(\ref{eq:x12}). The tail at larger $x$ is due to the smearing of the distribution arising from the ($qg\to \gamma^* q $)
subprocess. A narrow peak near the lowest possible value of $x$ is due to the $1/t$ singularity in the subprocess cross section (\ref{eq:A}). 

To calculate the distributions shown in Fig. \ref{fig:hist} we introduced the infrared cutoff $|t|>Q_0^2$, with $Q_0=1$ GeV (the smallest virtuality at which the DGLAP evolution starts). For a larger cutoff\footnote{In the case of 
 $Q_0\to 0$ we come back to $\delta$-functions at the $x$ values given by 
 (\ref{eq:x12}).}, $Q_0=2$ GeV
we get the lower peaks shown in the diagram. For the cutoff $Q_0=1$ GeV, the subprocess 
 $gq\to q\gamma^* $ gives more than 90\%  of the whole Drell-Yan cross section. Thus the 
 distributions shown in Fig. \ref{fig:hist} provide a rather good evaluation of 
the $x$-regions sampled by the Drell-Yan process for these kinematics.

What information about PDFs do we obtain from measurements of low-mass Drell-Yan lepton-pairs at high rapidity?  It is important to emphasize that the data make a {\it direct} measurement of the quark/antiquark PDFs at the optimal scale $\mu=1.4M$, with little factorization or renormalization scale ambiguity, for the values of $x_{1,2}$ given by (\ref{eq:x12}). That is, there is a direct one-to-one correspondence between the data and the PDFs.
For example, the production of a low-mass $M=6$ GeV $\mu^+\mu^-$ pair at 7 TeV, going forward with rapidity $Y=4$, measures quark/antiquark PDFs with
\begin{equation}
x_1=4.7 \times 10^{-2},~~~~~~x_2=1.6 \times 10^{-5},
\end{equation}
but at a moderately high scale, $\mu^2 \simeq 70 ~\GeV^2$.

What is the impact of such measurements on global PDF analyses? Clearly, the measurements can probe a small $x$ domain so far unexplored by data.  If, at such small $x$, we were to believe in pure DGLAP evolution, then the majority of the low $x$ quarks (antiquarks) comes
from the $g\to q\bar q$ splitting. Therefore the $M$ dependence of the Drell-Yan cross section will probe the gluon distribution in a similar way to how the scaling violations in deep inelastic scattering data,  
$dF_2(x,Q^2)/dQ^2$, probe, via DGLAP evolution, the low $x$ gluon density\footnote{Recall, that starting the evolution from $Q_0=1$ GeV more than 90\% of the Drell-Yan cross section is described by the 
$gq\to\gamma^*q$ subprocess, and so will probe $g(x_2)q(x_1)$, see Fig. 4.}. So, by including  low-mass high-rapidity Drell-Yan data in the global parton analysis we will strongly constrain the low $x$ gluon PDF. Note that Drell-Yan data at the LHC may probe very low values of $x$.  In this future global analysis 
 we should account for the absorptive, ln$(1/x)$, etc., 
 modifications to the evolution equations, as well as using the 
 Drell-Yan data at the optimal scale.

\begin{figure} [htb]
\begin{center}
\includegraphics[height=8cm]{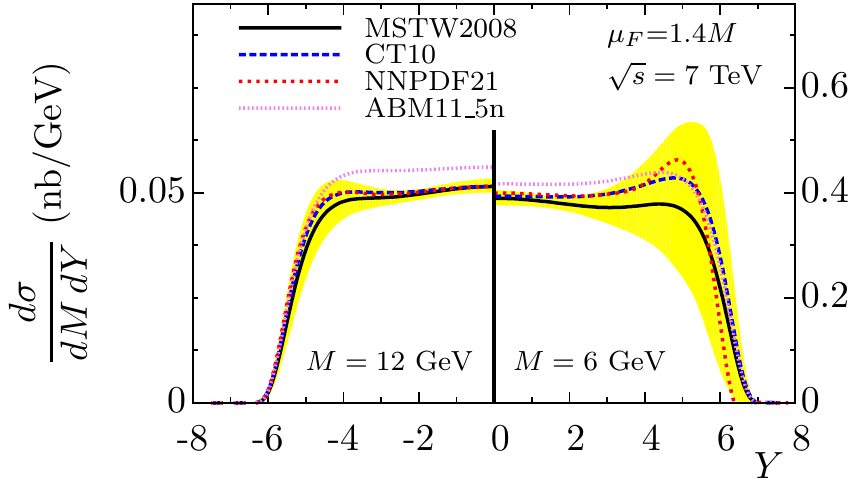}
\vspace{-0.2cm}
\caption{\sf The 'LO+NLO' cross sections for the Drell-Yan production of a $M=6$ and $M=12$ GeV $\mu^+\mu^-$ pair, as a function of its rapidity, at 7 TeV, obtained using four different recent NLO sets of PDFs \cite{MSTW, CT10, NNPDF21, ABM11}. We show the 1$\sigma$ error corridor for the predictions obtained using the MSTW parton set.}
\label{fig:pdfs}
\end{center}
\end{figure}
Bearing in mind the above qualifications, let us use recent sets of existing global PDFs to calculate the cross section for the examples of $M=6$ and $M=12$ GeV lepton-pair production. The results are shown in Fig. \ref{fig:pdfs}.  The scales in the figure have been adjusted to allow  for the naive $1/M^3$ behaviour of the cross section, so that any departure from the `symmetry' of the plot is due to the PDFs.

At first sight, it might be surprising that there is not more difference between the predictions from the different sets of PDFs for such low values of $x_2$. However, 
\begin{itemize}
\item[(a)] the predictions made by using all {\em reasonable} sets of PDFs which are consistent with each other within the error bands;
\item[(b)] the partons are being probed at a moderately high scale: $\mu^2_F\simeq 70$ and 280 GeV$^2$ where the difference between different sets of PDFs is not large;
\item[(c)] recall that, at the moment, there is {\em no} prediction for $Y\gapproxeq 4$, but only those based on `ad hoc'  
PDF extrapolations to very low $x$; however, the forthcoming LHCb data can provide an important direct {\it measurement} of the  PDFs in this low-$x$ region; a domain which is not accessible by previous data.
\end{itemize}


\section{Conclusion}

The observation of low-mass Drell-Yan $\mu^+\mu^-$ pairs at high rapidity at the LHC can probe the  PDFs in a low $x$ domain unreachable in previous experiments; that is, $x \lapproxeq 10^{-4}$.  However, the constraints on the PDFs found using a conventional analysis of the data, show great sensitivity to the choice of factorization (and renormalization) scale. Here we have introduced a technique to determine an optimal scale which greatly reduces this sensitivity. The optimal scale is found to be $\mu_F=1.4M$, where $M$ is the mass of the produced $\mu^+\mu^-$ pair.  Simply for illustration, we show `predictions' of the  Drell-Yan cross section, as a function of rapidity, $Y$, for $M=6$ and $M=12$ GeV for a variety of recent PDF sets, but we stress that the predictions for $Y \gapproxeq 4$ depend on `ad hoc' extrapolations. Here, LHC data can, for the first time, make a direct measurement of PDFs in the $x \lapproxeq 10^{-4}$ domain.

\section*{Acknowledgements}
We thank Ronan McNulty for useful discussions. EGdO and MGR thank the IPPP at the University of Durham for hospitality. This work was supported by the grant RFBR 11-02-00120-a
and by the Federal Program of the Russian State RSGSS-65751.2010.2;
and by FAPESP (Brazil) under contract 2011/50597-8.

\end{document}